# Mechanisms for producing a working knowledge: enacting, orchestrating and organizing

Gunnar Ellingsen and Eric Monteiro

gunnar.ellingsen@unn.no and eric.monteiro@idi.ntnu.no

Dept. of Computer and Information Science

Norwegian University of Science and Technology

N-7491 Trondheim

# Mechanisms for producing a working knowledge: enacting, orchestrating and organizing

**Abstract.** Given that knowledge (intensive) work takes place immersed in truly heterogenous networks of knowledge representations (codified, narrative, embedded in routines, inscribed in artefacts), our analysis is geared towards how the transformation of these resources are enacted in the practise of everyday, knowledge work. First, we discuss the work, strategies and mechanisms implied in rendering knowledge as credible, trustworthy and relevant. Second, we analyse how sediments of historically superimposed layers of knowledge representations need to be enacted through selective repetitions, omittance and highlighting to preserve it as 'living' knowledge. Third, supplementing the more cognitivelly oriented aspects of knowledge work, we discuss how codified knowledge representations organise, coordinate and delegate work. Empirically, we study clinical work in large hospitals, a type of work, we argue, that unduely has been left out of traditional listings of knowledge work.

# 1. Introduction

The establishment of knowledge (intensive) work as a research theme has spawned considerable interest into characterisations, typologies and underlying conditions for this kind of work (Zack, 1999; Blackler, 1995; Nonaka and Takeuchi, 1995; von Krogh, 1998). Yet, as Alvesson (2001) so timely reminds us, this does not avoid the distinct bias in much of the approaches, definitions and conceptualisations of knowledge work as they are 'likely to be contestable' (Alvesson, 2001:864).

The preoccupation, bordering on obsession in knowledge management research, with engineers, designers and consultants needs to be recognised as ideologically rather than analytically founded. Sturdy (1997), for instance, presents a picture of consultants and consultancy work as routinised and haunted by anxieties about staging a facade of expertise both to clients as well as to colleagues. It is neither obvious nor 'natural' that characteristics of knowledge work as specialised, requiring a high level of formal training and innovativness translates into the kind of work listed earlier. To take but one example, Knorr-Cetina (1999) argues convincingly that the organisation of work within scientific laboratories – through increasing, global competition currently under intense pressure to transform into more collaborative coalitions - might very well function as configurations of knowledge work more in general (see also Boland & Tenkasi, 1995).

This apparent lack of precision in the conceptualisation of what constitutes knowledge work does not, however, imply that the term is void for further analysis as "it makes sense to refer to knowledge-intensive companies as a vague but meaningful category" (Alvesson, 2001:864). It is all too easy to entangled in extensive elaborations on definitions, conceptualisations and perspectives on 'knowledge'. Here, we employ a working definition of knowledge as the action-oriented capacity based on – but not reducible to – various forms of knowledge representations (such as artefacts, written and oral information). The thrust of our analysis focuses on the process of how this extensive and heterogeneous network of knowledge representations is transformed, translated and moulded into working knowledge (Orlikowski, 2002).

The study reported here focuses on clinical work (diagnosing, treating, curing and checking patients) within large hospitals. For the reasons indicated above, the clinical work of physicians in hospitals tends to get bypassed in most discussions on knowledge work. This is unfortunate as clinical work – highly influenced by the scientifically legitimised knowledge production, involving highly educated personal, characterised by interdisciplinarity, subject to high degrees of risk – provides an instructive occasion to study knowledge work in action.

There is clearly a cognitive complexity to clinical practise including puzzle solving over diagnosing and keeping updated on recent research progress and medications. Still, the complexity and aspect of knowledge work we emphasise here is linked to the intrinsically *distributed* character of clinical work as "a diagnosis … can be not only cognitively, but also socially complex" (Cicourel, 1990:222). There is, quite literally, no single individual who possesses the complete knowledge about any given patient. Rather, it is dispersed throughout a truly vast, heterogeneous network of individuals, communities, archives and information systems (Berg, 1996, 1999; Strauss et al., 1985; Atkinson, 1995). This observation marks the point of departure of our analysis that subsequently evolves along three themes.

First, the *enactment* of the various elements in the heterogeneous network go well beyond a mere 'collecting' of given elements. Working knowledge is moulded and crafted through re-presentations, thus creating a genealogy of historical sediments of layers of knowledge representations (Orlikowski, 2002). Despite the apparent repetitious nature of the 'gathering' of knowledge representations, the crucial task in knowledge work is to re-vitalise them by enacting them. A crucial aspect of this process, which more often than not tend to be glossed over in accounts of knowledge work, is the preservation of earlier accounts while at the same time adding new layers, new versions. In keeping up with the vast, accumulated body of knowledge representations of a patient (history of hospitalisation, accounts from other departments and laboratories), it is crucial to craft, mould or narrate a more manageable trajectory which also serves as an 'organisational memory'. These summaries act, to use Bowker's (2002) phrase, as 'folded histories' as they selectively enact historically buried representations of knowledge.

Second, the bringing together or *orchestrating* of the many forms, sources and representations of knowledge is analysed. Moving beyond simplistic dichotomies like, say, tacit/ explicit knowledge representations, we dwell on the practices for rendering knowledge (regardless of representation) credible, relevant and trustworthy. It is not immediately obvious which sources of knowledge are credible and which are less so (Shapin, 1994). Sorting this out involves work like double checking by looking the same information up in an alternative source, discussing it with members of your community or relying on your existing, social network. The working knowledge emerges through a fluid interleaving of the various sources and forms of knowledge representations; in essence, knowledge creation is orchestrated.

Third, the knowledge representations also play productive roles beyond feeding into the knowledge work itself. They are key vehicles in the actual *organisation* of the work as they coordinate, delegate and distribute work across time and space and professional groups (communities) (Berg, 1996, 1999; Hutchins, 1994; Smith, 1990). How knowledge work actually gets organised is a downplayed aspect of the complexities of this kind of work at the expense of the intellectual (Boland & Tenkasi, 1995), social networks (Blackler, 1995; Lave and Wenger, 1991; Brown & Duguid, 1991; Nonaka & Tacheuchi, 1995) or emotional (von Krogh, Ichijo & Nonaka, 2000) aspects. Beyond pure knowledge representations, utterly mundane artefacts such as forms, sheets and reports simultaneously function as tokens signifying the completion of one task and the handing over of the responsibility to someone else.

Empirically, we analyse so-called discharge letters. Their production provides a particularly relevant instance of the themes outlined above. The discharge letters are worked out by the physicians upon the departure of the patient from the hospital. Rather than 'collecting' the 'facts' of what happened during the patient's stay, it is an occasion for enacting and orchestrating the distributed knowledge representations, crafting it into a narrative useful for its recipients who predominantly are physicians in primary health care (the patient's general practitioner) or physicians at the patient's local hospital. In addition, the hospital physicians themselves use them whenever the patients return to the hospitals. They accordingly are the topmost layer of the genealogy of knowledge representations that in sum make up what 'the hospital' knows about a patient; they are the invoked remembrance of a patient's trajectory.

In section 2 we outline our framework for conceptualising knowledge work. We emphasise its narrative form, the heterogeneity and the historicity of knowledge representations. Section 3 outlines the setting of the study at the University hospital of Northern Norway (UNN). It also discusses methodological issues. Section 4 contains four case vignettes from different wards at the hospital. The analysis is contained in section 5 and is structured as indicated above into the enactment, the orchestrating and the organisation of knowledge work.  Section 6 concludes by pointing out how our analysis makes visible largely overlooked (in the dominant positions on knowledge work) activities intrinsic to knowledge work. We raise, without resolving, the question of what role information systems could and should play given our account of knowledge work, including addressing the issue of integration of information systems.

## 2. Conceptualising knowledge work

The deeply social aspects are a pronounced theme in much of the writings on knowledge work. It is, rightly so, pointed out how social networks in general and communities of practices in particular capture an important way in which knowledge is formed, created and circulated. Learning and knowledge sharing does not take place isolated from or 'above' social interaction; it is an intrinsic part of the constitution of communities (Lave & Wenger, 1991; Orr, 1995; Nonaka & Takeuchi, 1995). One aspect of how knowledge circulates within a community of practice is the emphasis, inspired in part by Bruner's (1986) arguments, of the narrative format of knowledge. Thus, knowledge workers are involved in learning and creation of new knowledge by becoming 'insiders' in the community (Brown & Duguid, 1991:48), that is, they are acquiring not explicit, formal "expert knowledge, but the embodied ability to behave as community members". This strand of studies of knowledge work explores how narratives are formed and how they operate in communities (see Orr, 1995; Boland & Tenkasi, 1995, Brown and Duguid 1991; Czarniawska, 1997). Within health care, the importance of narratives has also been emphasised (Atkinson, 1995; Hunter, 1991). Hunter (1991: 69) underscores the intimate and irreducible nature of both written and oral knowledge:

> "Physicians refer to written materials in the production of their spoken performances; the latter may provide the basis for subsequent written texts… a sharp distinction between 'literate' and 'oral' aspects of culture is misplaced in many instances: it is undeniably true of the culture of medicine"

The focus on narratives, taken to the extreme, may encourage a portrait of knowledge as essentially social and irreducibly by coding thus being akin to one of the positions suggested by Lam (1997:979):

> "Knowledge is generated and stored 'organically' in team relationships and the mode of coordination is human-network based. This type of knowledge is not amenable to systematic codification and can only be accessed and transferred through intimate social interaction"

This downplays to the level of non-existence the role of externalised or codified representations of knowledge, a position which ultimately is unattainable (Nonaka & Takeuchi, 1995).

The key issue, then, is to explore how a network of heterogeneous representations of knowledge gets orchestrated or brought together to a (reasonably coherent) body of working knowledge. This involves, as Boland and Tenkasi (1995:359) point out, an element of validating or sense-making of the different elements:

> "In summary then, the problem of integration of knowledge in knowledge-intensive firms is not a problem of simply combining, sharing or making data commonly available. It is a problem of perspective taking in which the unique thought worlds of different communities of knowing are made visible and accessible to others"

The dichotomous distinction heralded by Nonaka and Takeuchi (1995) between so-called tacit and explicit knowledge sidesteps the key issue of describing the work, efforts and strategies of rendering knowledge understandable, credible and trustworthy. Large organisations, including large hospitals, need to cope with what Giddens (1991: 21) argues to be an increasingly important aspect of the modern world and describes as 'disembedding', that is, "the 'lifting out' of social relations from local contexts of interaction and their restructuring across indefinite spans of time-space". An essential aspect of this is how processes of 're-embedding' need to re-establish social links, networks and communities of practise.

As convincingly demonstrated in later studies in knowledge management, externalised materialisations of knowledge (likely to face 'disembedding') require a social embedding (Lam 1997; von Krogh, Ichijo & Nonaka, 2000). The question, then, is how do organisations preserve the deeply social aspects of knowledge production under the increasing pressure of disembedding processes? On Giddens' (1991: 79-80) account, modern societies and organisations have to increasingly trust 'abstract systems' implying 'faceless commitments'. Hence, this delegates a pivotal role to establishing, maintaining and extending trust as a necessary basis for knowledge work and knowledge cultivation as argued by Shapin (1994).

The kind of trust involved in knowledge work is not a static entity either present or absent. It is rather the performed achievement of a concerted and highly heterogeneous effort with actors, artefacts and other externalised knowledge representations. As pointed out by Cicourel (1990:222), "the perceived value of medical information is related to the perceived credibility of the source". An important aspect of knowledge work, then, is to unpack how disembedded or externalised knowledge representations are rendered credible and trustworthy. This problem is dramatically reinforced in settings, including modern hospitals, with extensive generation of externalised knowledge representations.

Large hospitals depend heavily on textual practices to co-ordinate, monitor and organise relations between different segments and phases of medical work (Smith, 1990:217-218). This underscores how text is never neutral, a point also emphasised by Atkinson (1995:127):

> "Not all knowledge is treated as having equal value. It has different sources, and has different weight attached to it, and may be regarded as more or less warranted (…) in expressing his or her attitude towards facts and opinions, the clinician also inscribes aspects of the moral and technical division of labour among medical specialist".

Documents or texts are visible constituents of social relations (Smith, 1990:210). Texts contribute to "externalise social consciousness in social practices, objectifying reasoning, knowledge, memory (...) decision-making etc. as properties of formal organisation" as Smith (1990:210) points out before continuing:

> "The simple properties of the documentary or textually mediated forms of social organisation involve their dependence upon, and exploitation of, the textual capacity to crystallise and preserve a definite form of words detached from their local historicity"

In the process of drawing this vast network of knowledge representations together – narratives, patients records, laboratory reports, notes – there is an ongoing effort of highlighting some elements at the expense of others. Or as Bowker (1997:15) puts it by paraphrasing Garfinkel's classic paper, "I have argued that there may indeed be good organizational reasons for forgetting". This points to the essential, but less developed, theme within studies of knowledge work, namely the layering or *genealogy* of knowledge representations. Bowker (2002:5 (emphasis added)) makes the relevant remark that "[t]here has been relatively little work in … dealing with the organizational, political and scientific *layering* of data structures". Truly vast bodies of knowledge representations cannot, quite literally, be maintained in full. Through ongoing, selective forgetting and re-presentations, extensive bodies of knowledge representations are moulded into working knowledge. Bannon and Kuutti (1996) make the timely observation vis-à-vis the problematic notion of 'organisational memory':

> "if 'organizational memory' is at all a useful concept, it is so to the extent that it refers to active remembering which carries with it its own context – so that it comes in the form not of true or false but of multifaceted stories open to interpretation"

The layering or genealogy of knowledge representations implies that these representations need to embed some kind of historical context although "[w]e cannot retain everything about a set of data (this would be bureaucracy gone wild)…[we need] historical perceptions of data" (Bowker, 2002:33).

The amount of codified knowledge representations in hospitals is significant. To illustrate, the paper based patient records at one of these hospitals (RiT Trondheim, Norway) occupies 16 km of shelves. The specific task of physicians we focus empirically on in our study is the writing of discharge letters. These letters are to bring together the truly dispersed and heterogeneous knowledge representations of what 'happened' during a patient's stay. Thus, the re-telling and enactment involved in writing the discharge letters goes well beyond a mere recombination of existing 'facts'. It adds an interpretation and is crafted with a purpose in mind. In writing these discharge letters the physicians simultaneously add to the existing layers of historical accounts of the patient's trajectory. A key recipient group is the general practitioners, a community of practice distinct from the hospital based physicians who produce the discharge letters. This implies that their production involves the kind of perspective taking that Boland and Tenkasi (1995) emphasise.

Berg (1996:5) reminds us that the mode in which clinical work takes place is geared towards 'what to do next':

> "Through [the physician's] activities of reading and writing (…) he narrows down the plethora of potential tasks and divergent data into a clear notion of 'what to do next'"

One aspect of the 'what to do next' framing of knowledge work for hospital physicians, is the way externalised representations of knowledge – forms, reports, records – at the same time function as cues or tokens that feed into the coordination, delegation and accountability of the work, also of nurses, secretaries, physiotherapists and other professions (communities) at a hospital (Berg, 1999; Smith, 1990). Hence, externalised knowledge representations also play a performative role in the everyday organisation of hospital work in total, an organisational complexity that exceeds any individual's capacity (Hutchins, 1994). Knowledge work is thus not 'pure', that is, independent of and above the more mundane task of the organisation of everyday work.

## 3. The setting of the study

This study belongs to an interpretative approach to the development and use of information systems (Klein & Myers, 1999; Walsham, 1993).

The study has been conducted at the University Hospital of Northern Norway (UNN). The hospital has approximately 4000 employees, including 400 physicians and 900 nurses. The hospital has 600 beds of which 450 are somatic and 150 are psychiatric. We rely on four types of data: participative observations, interviews, informal discussions and electronic and paper-based documents. The participative observations took place during January-March 2001 at 4 of the hospital's clinical departments. In total, 42 hours were spent observing work. In addition, 34 semi-unstructured interviews where conducted during the periods mentioned above. Each interview lasted from 1-2 hours.

To further discuss our methodological approach, we relate it to the seven principles of Klein and Myers (1999).

The first is the 'fundamental principle of the hermeneutic circle' which underscores that insight into a complex whole is gained through meanings about its parts and their interrelationships. This amount to making sense of events and episodes against a broader backdrop of an understanding of the whole. For us, this was due to a pre-understanding stemming from the authors' existing relations and experience with the health care sector over the last ten years (see details further below).

The second principle is that of 'contextualization'. We take this principle into account by acknowledging that we for some time have being doing research in this area. In addition to the data that is a direct foundation for this paper, the first author has conducted 57 hours of observation in four other departments. Also, our ten years engagement as researchers and practitioner in the field gave us a sense of the historical development underlying the current initiatives.

The third principle of 'interaction' between the researchers "requires critical reflection on how the research materials ('or data') were socially constructed through the interaction between the researcher and the participants" (Klein and Myers, 1999:72). During the observations, the first author was recognised as already having a role. Before embarking on a research career, he had been employed as an IT-consultant at UNN for 10 years. Hence, informants occasionally challenged him about how to design EPRs for physicians. This made us also reflect on how we could contribute with practical design guidelines for IT-systems in hospitals. We have received half a A4 page of written feedback on email to an earlier draft of this paper.

In order to make the observations as smoothly as possible, the first author was wearing a physicians' white coat. The experiences with – and without – this coat were quite remarkable as it ensured community membership. People did not seem to bother about being observed, due in part to the use of the white coat. This resulted in a fluctuation between a fairly passive role merely observing as non-obtrusively as possible and a more active role, when possible, posing questions for clarification and explanation.

Our understanding of the material has also been shaped by discussions with health workers at seminars, meetings and informal discussions. The authors have run over thirty seminars over a 3-year period where we have invited guests (practitioners, designers and researchers) who have presented relevant themes within this area, thus providing feedback on our interpretations.

The fourth principle of 'abstraction and generalisation' is accounted for in the introduction of this paper where we outline how we approach the concept of knowledge.

The fifth principal of dialogical reasoning accounts for our development of how we finally have chosen to present the case to readers of this paper. As a starting point, we adhered to a 'realistic' style, primarily out of habit. A realistic style focuses on thoroughly mundane details of everyday life such as regular and often-observed activities of the group under study (van Maanen, 1988:48; Schultz, 2000). However as the data emerged, we were increasingly inspired to frame our data in a 'impressionist' style characterised by 'fleeting moments of fieldwork cast in dramatic form (van Maanen, 1988:7) as it enabled and 'allowed' us to focus on particularly interesting aspects of the data.

We have taken the sixth principle of multiple interpretations into account by letting several kinds of informants speak. Even if the observations where especially aimed at work situations for physicians, secretaries as well as nurses were partly involved. On occasions, also patient examinations were observed. This induced various perspectives on meanings about similar issues. Posted questions were also part of the observations in order to clarify and elaborate observations. The extent and format of these obviously varied with what was possible without intruding too much with ongoing work. For instance, questions were postponed when the work was recognised as hectic, during formal group meetings or in front of patients.

## 4. Four case vignettes

Medical practice varies enormously – within different domains, departments, hospitals and countries (Atkinson, 1995; Strauss et al., 1985; Berg, 1999). We have no ambition of paying justice to this variation in any systematic or comprehensive manner. Rather, our aim is to motivate for an appreciation of this variation through a sampling of four wards at University hospital of Northern Norway (UNN). This variation in practice also translates into a corresponding variation in the generation and use of representations of knowledge about the patient. The observations are especially targeted at the process of producing discharge letters. Characteristic features of the work situation in the wards are

1) *Dept. of Ear, Nose and Throat*: A small, largely self-contained surgical department where the patient have relatively precise diagnoses.

2) *Dept. of Cardiac and Thoracic Surgery*: An extremely hectic, highly specialised department with relatively narrow problems of concern. Most of the patients have had a full investigation in another department.

3) *Section of Nephrology, Dept of Medicine*: A section dominated by patients with chronic diseases. As a part of the Dept of Medicine, they receive a fair share of undiagnosed emergency cases.

4) *Dept. of Oncology*: Most of the daily operations of the department are scheduled but due to complex diagnoses, the physicians have to deal with a high degree of uncertainty.

### 4.1. Department of Ear, Nose and Throat – business as usual

Dept. of Ear, Nose and Throat is a ward dominated by many small surgical operations. This makes the cases relatively predictable. In order to ensure efficiency in the production of discharge letters, the department routinely reuse documentation from the patient record. We illustrate the work at the ward:

> The examination room has the ascetic, non-personal appearance of most examination rooms. Colouring is discrete and lighting brisk. The apparent untidiness of the large desk with piles of documents, notebooks, short lists and one computer signal a state of alert. The chief physician and a nurse, wearing their white frocks, are preparing for the examination of six patients. The examination chair – excessively equipped for flexible handling – dominates the room like a sculpture. It resembles a dentist's chair.

The patients are transient and about to leave the hospital. The patients are admitted into the room one at a time and requested to mount the examination chair. The chief physician immediately and rapidly examines their throats and noses. There are few detours. A typical examination lasts a couple of minutes. The patients' general health condition is good. For instance, a young patient, hospitalised due to complications after surgery to his tonsils, receives a discharge letter stating "…beyond that, the patient is in good health".

The dialogue between the health personnel and the patients is relaxed and sprinkled with jokes. While cleaning an old man's nose, the chief physician refers to the patient's librarian duties by jokingly instructing him to "stay away from those dusty books". Even if the chief physician and the nurse are under pressure to finish off awaiting patients, the atmosphere remains calm. The computer is not used during the examination as the chief physician has a clear sense of the patient's condition. Occasionally, he needs to look up x-ray reports and laboratory results from the paper-based patient record and the various information systems.

Before the next patient is admitted, the chief physician dictates the discharge letter for the last patient. More specifically, he only dictates the last part of it, the conclusion. This takes less than one minute and amounts to five lines of text. This conclusion contains diagnosis and procedure codes drawn from a short list of regularly used diagnosis codes posted at the desk.

The rest of the discharge letter is reused from existing documentation already produced during the stay. To implement this reuse of text, the chief physician ticks off a box on the paper form that accompanies the dictated tape. This is then handed over to the secretary to transcribe and edit before the full text version is returned to the chief physician for final signing. .

## 4.2. Department of Cardiac and Thoracic Surgery – interleaving the oral and the written

The Dept. of Cardiac and Thoracic Surgery is responsible for cardiac surgery for adults in the Northern Health Region of Norway as well as regionally responsible for general thoracic surgery. Most of the patients have already received a full examination by another department or (local) hospital. The patients normally stay for about a week. After their surgery, they are transferred back to another department within the hospital or to a local hospital. The discharge letter is the key vehicle for communicating to the recipients the relevant insights gained and instructions for further follow-up. The following vignette illustrates the work in the department:

> The ritual beginning of a new day at the department is the morning meeting. A head physician, three assistant physicians and six nurses are seated around the table dominating the room. In a focused, high-paced rhythm the patients are discussed in turn. The discussions circle around the heart surgery; prescribed medication, further treatments and whether the patient may leave today or not. The nurses come and go continuously. They stay for 'their' patients but skip the others'.
>
> We follow one of the assistant physicians, Pasi, who leaves the morning meeting to produce a discharge letter for a patient leaving later this very day. At the meeting, he was informed about the current status of the patient. He retrieves the paper-based patient record from the on-duty room and he brings it to a desk. From the bulky patient record, Pasi first selects the admittance report and the surgery report from the previous day. He also digs out and reads the discharge letter from an earlier stay at the Dept. of Medicine one month ago where the patient had a full examination prior to his surgery. He takes his time reading, maybe 15-20 minutes. "The patient's kidney was removed in 1964 caused by kidney stones", Pasi comments, "That's very unusual! Perhaps it was a big one?".

The on-duty room is busy and crowded. Physicians and nurses pass through all the time contributing to a hum of questions, advice and discussions. The daily operations at the ward are listed on a big board occupying one of the walls. The phone cannot keep quiet. Ignoring the surrounding noise, this is where Pasi usually dictates the discharge letters. Until the patient is discharged, the paper-based patient record, the chart book and the nurses' documentation are located somewhere in the ward, normally right here in the on-duty room. Staying here makes it convenient to ask for additional information, as the patient is still fresh in the minds of everyone. Pasi starts dictating the social status of a patient who had heart surgery: "72 years old fisherman that lives together with his wife…". The letter is concise and terse in style. As one of the head physicians puts it: "It has to be short to get to the point". In his dictation, Pasi does not reuse the summary from the admittance report (as some do) as: "I use it if it is good, but not always", before elaborating "most of my time I spend obtaining an overview of the case. For this I have to read in the patient record. It might be that documentation produced some time ago is important. Also, reports from surgical and medical wards are very different [in style], which means that sometimes I have to turn to old documentation to get the whole picture". As a relatively inexperienced assistant physician at this department, he often has to spend 2 hours reading to get an overview of a case. His dictation is rather staccato as he tries to make sense of information from several sources: both from the pile of papers from the patient records as well as from the electronic patient record. At one point, he realises that he needs the patient's chart. He stops dictating, enters one of the examination rooms, locates the chart and continuous to dictate. He also reads and selects information from the nurses' report. Once more, he stops dictating, picks up some medical measuring device and visits the patient to measure her talus arm index. Shortly after, he returns to his desk commenting "no pulse in the foot, there is better circulation in the minor arteries, but not in the large ones". As a final but vital point, he has to decide whether the patient needs to be summoned to a later control. For this, the chief physician needs to be consulted. Pasi leaves the room, searches for him, locates him in one of the patient rooms and obtains the necessary information to finish his dictation.

It is time to find a secretary who can type what he has dictated. Pasi hurries around and looks for a particular one. After he has found her, she gets the tape and starts to write immediately. Pasi goes back to the on-duty room, waits by the computer, ready to proofread and sign the discharge letter as soon as the secretary has finished typing. In the mean time, the nurse responsible for the discharge of the patient enters the room and asks for the report. She is put under a certain strain since the ambulance air transport is ready to take off and leave for the local hospital and they await the report. Now both Pasi and the nurse are waiting and in couple of minutes it appears. Pasi proofreads it and signs it before handing it over.

## 4.3. Section of Nephrology, Dept. of Medicine – maintaining a working memory

The Section of Nephrology is a part of the Dept. of Medicine. The section has many patients related to chronic diseases (such as kidney failure) who come for periodic treatments. In addition, as a part of Dept. of Medicine, the section has to respond to emergency patients with undiagnosed problems. An illustration of the work of an experienced physician is presented below:

The physician has just managed to break away from the daily buzz of patient-related work to produce discharge letters. The patients involved have been discharged from the hospital a couple of days ago, but this is his first opportunity to finish off work related to their departure. He is able to find a vacant office at the ward where he brings his pile of paper-based patient records. Obviously, the others at the ward know where he is as they pop into the room to make inquiries. He also has to respond to his beeper, but this does not force him to leave the room.

"RETURN DIALYSE" is written in large letters on the front covers of several of the paper records. This means that these paper records are stored in the Peritoneal Dialysis (PD)-section in a special archive. They belong to a special type of patients who come regularly. As a result, only the secretaries in the PD-section write these reports in order to ensure that everything is done right.

Labouring through his workload, it becomes clear the several of the cases are quite complex: unscheduled emergency cases, several examinations in other departments, contacts with psychiatric sections or dependence on results from several laboratories. He keeps pausing to check information from several sources such as blood results from the laboratory system as well as running notes, results from referrals and patient charts from paper records. Together they constitute pieces in a puzzle that need to be assembled, evaluated and assessed closely with the patient's current condition.

He reads extensively. Reading and dictating are interleaved. The physician also needs to consult a couple of his colleges by phone.

For some of the patients, he accesses the electronic x-ray system. He examines both the x-ray pictures and the x-ray reports before making a summary of it on the fly while dictating. He also includes his own assessments. Retrospectively explaining this, he states that "Sometimes I can cut and paste from parts of the x-ray report. It depends on how much of it is important".

One of the discharge letters is based on an emergency admission. This time he dictates, in part, the same information that is in the admittance report. He explains that the reason why he did not instruct the secretary to reuse the first sections (cf. section 4.2) was because the information in the admittance report was incorrect. He had to correct this information based on conversations with the patient and his wife. He also studies the nurses' reports and says: "You have to do that often to check whether it contains something important", before concluding for this patient, "well …nothing important this time". For the next patient, however, the physician instructs the secretary to copy from named sections of the admittance report. The physician explains afterwards that he knew what the documentation contained because he had dictated this admittance report himself.

While dictating procedure codes, he exclaims: "This is not correct!" and starts the code tool (a dictionary) on the computer. It is used to find the right procedure- and diagnose-codes (NCSP and ICD[1]). After a while he finds the proper codes and dictates them, sighing: "This coding is definitely what consumes most time when dictating discharge letters".

The final patient is a chronic Peritoneal Dialysis patient. In addition to the dictation, he retrieves the patient's Peritoneal Dialysis form (see Figure 1) from the computer. He copies it and pastes it into the discharge letter. It contains a lot of important measurements related to the patient's condition. It acts as a working 'memory' as: "This is a patient that regularly returns to the section and he needs clear cut rules for who is responsible for what. As is possible to see here [pointing ], PET analysis is not performed during this stay, but down here [pointing to the bottom of the form] you can see that it is decided that it will be carried out during the next stay". When Peritoneal Dialysis patients are hospitalised, it is standard procedure to check the most recent discharge letter to see whether any special tests are planned.

| Date | 10.01.01 |
|---|---|
| Estimated dry weight | 87-88 kg |
| weight | 87-88 kg |
| Quantity of urine | 2000 ml |
| Quantity of dialyse solution | 12730 ml |
| Ultra filtrate | 1000 – 1400 ml |
| Blood pressure | 172/106 |

---

[1] NCSP is an abbreviation for NOMESCO Classification of Surgical Procedures. The NOrdic MEdico-Statistical Committee was set up in 1966, following a recommendation by the Nordic Council. An aim of NOMESCO is to promote the coordination of health statistics in the Nordic countries. International Classification of Diseases (ICD) is worked out by the World Health Organization (WHO).

| KT/V –own | 1,16 |
| --- | --- |
| -total | 3,09 |
| Kreatin-clearance | 74,6 l/week |
| PET | Not performed at this stay |
| Prealbumin/albumin | 35 |
| HB | 12,6 |
| iron | 17 |
| TIBC | 62 |
| ferritin | 205 |
| Ca /ion. Calcium | 2,36/1,22 |
| Phosphate | 1,92 |
| PTH | 20,4 |
| Bag strength | CAPD bag strength Locolys 2,3 % 4 x 2 liter at day time, 2,5 liter Extraneal at night time |
| Exit-site | Good |
| Next control | 1. At Medical policlinic. With MRU, 2 months. With measuring of rest function. 2. New PD-control with PET in May-02 |
| Signature | NN |

**Figure 1 The Peritoneal Dialysis (PD) form, which forms a part of the discharge letter**

### 4.4. The Dept of Oncology –narrative sense-making

The Dept. of Oncology is the only one in Northern Norway and is the principal hub for cancer treatment in the region. Most of the patients are examined at a local hospital before they are admitted to this department. Due to the character of the disease, some of the patients are hospitalised for a relatively long period of time. Others, depending on the treatment protocols, receive periodical treatment such as radiation or chemotherapy treatments. The bulk of the daily activities are pre-scheduled. Yet, the work tends to be hectic as many of the patients have complex trajectories, which make time estimates strewn with errors. We follow an assistant physician with four months of experience:

> We are in the assistant physician's office outside the ward. It is six o'clock in the afternoon. It is quiet. The room is semi-lit. The computer screen beams brightly at the desk. The daytime workload makes discharge letter production prohibitive. They end up as evening activities and are dictated in her office. The physician is relaxed, neither the beeper nor the phone interrupts the work and at regular intervals she allows time to explain what she is doing.
>
> The discharge letters from this department are extensive. There is little given structure to the text; they come close to free-text descriptions. The first patient is a newcomer, requiring extra time to study. She spends some time going through the paper record as well as retrieving information from the electronic patient record. She dictates extensively, describing the current situation for the patient. She logs onto the electronic x-ray system and reads the CT-description, makes a summary of it and continues to dictate. Part of her dictation shows that during the stay the physicians have discussed possible treatment alternatives. The patient has also been involved in these discussions and has insisted on trying a special treatment that the patient has become aware of may have effect on his diagnosis D1. The patient is now discharged from the department without any documented effect of the current treatment.
>
> After the stay, the physicians have continued to evaluate the case and have agreed to invite the patient to an ongoing research study outside the given protocol. This implies that the patient will receive treatment T1. A complication is that the patient may also suffer from diagnosis D2. If so, normal treatment (T2) would have been given, but the physicians have never previously combined treatment T1 and T2. One of the involved head physicians recommends that they order treatment T1 (and not treatment T2) even if the patient also suffers from D2. The assistant physician works over half an hour with this discharge letter, producing two pages of text. Afterwards she explains that they treat many different types of cancer, some of which are rare for general practitioners. It is accordingly difficult to know what to be aware of, for instance possible side effects of certain treatments. These are rather specialised things that should be included in the discharge letters.

# 5. Analysis

## 5.1. Enacting

The importance of narrative forms of knowledge is already well rehearsed (Orr, 1990; Czarniawska, 1997; Bruner, 1986; Boland & Tenkasi, 1995; Brown & Duguid, 1991). In the present context of clinical work with its heavy reliance on the truly vast amount of codified knowledge representations (cf. section 2), the narratives function as means for rendering knowledge credible. The narrative is closely interleaved and operates in tandem with, rather than substituting for, codified representations. As Atkinson (1995:91) points out:

> "There is in the everyday organisation of medical work a close relationship between written and oral accounts constructed by medical practitioners for their colleagues"

In this understanding of medical work, working knowledge is enacted within an oral culture and the medical work is constantly produced and reproduced through a 'narrative encapsulation' of the knowledge (Knorr-Cetina 1999).

The close interleaving of the oral and the written is institutionalised by a number of rituals including clinical lectures, ward rounds, mortality and morbidity review and a surgeon's commentary to assistant physicians and students. More specifically, the role of narratives in clinical work is essential in rendering knowledge credible. Potentially disturbing to some perhaps, clinical expertise is neither absolute nor infallible. It is constantly subject to doubts, diverging views and negotiations (Hunter, 1991:28):

> "The epistemological importance of narrative is medicine's responses to the uncertainty inherent in its predicament as a science of individuals. Because the uncertainties of diagnosis and prognosis are fundamental to medicine, the methods physicians have devised to meet them are a fundamental part of medicine as well"

The Dept. of Rheumatology provides a vivid illustration of this. At the morning meetings, facing patient records each with 20cm of written documentation resulting from the chronic conditions, the narrative encapsulation is vital as one of the physicians explains:

> "We are a kind of oral and assessing profession (…) it is important to have meetings, to discuss which treatment that is most important or correct and whether it should change or not (...) [for chronic patients] we have medications that will not be effective within 3 months or 6 months time"

Narratives are of course not the only means of rendering knowledge representations credible.

A relatively downplayed theme in discussions on knowledge work in action is the way knowledge is historically stratified, not in any strict, accumulative fashion but rather as a genealogy of sediments. For a start, this implies that the history needs to be reconstructed and the discharge letters have a key role in this:

> "The discharge letter is the first document you will look into when establishing an overview of what happened with the patient during hospitalisation. If this is insufficient, you have to look into the running notes and if it still is not enough you have to look into the patient charts in order to check the medications that nonetheless should be reflected in the discharge letters. Eventually, you have to look into the nurses' reports" (physician, Section of Nephrology).

This is clearly illustrated when Pasi (Dept. of Cardiac and Thoracic Surgery, section 4.2) dictates the discharge letter, which summarises the patient's surgery. Note how he strategically selects the one-month old discharge letter from the Dept. of Medicine to learn about the case. As the Dept. of Medicine gave the patient full examination prior to the surgery at Dept. of Cardiac and Thoracic Surgery, this is hardly surprising. It illustrates, however, how the discharge letter goes beyond a purely 'accumulated' knowledge of the case. Rather, it represents the current sediment in the case trajectory. Enacting these historically buried sediments may prove essential in order to achieve sufficient insight into a case. As Pasi explains when recollecting his days as a newly hired assistant physician:

> "Most of the time, I spend getting an overview of the whole picture. Thus, I have to read, as it is difficult to know beforehand what is relevant. It could be that old information proves important, despite the fact that this information would not be summarised in the last discharged letter. In my early days, I sometimes spent 2 hours reading to gain an overview of a case"

A key insight is the way *repetitions* carry weight; they are anything but void. Repetitions selectively enact certain elements by omitting others. The generation of discharge letters, intended to summarise and 'repeat', provides an opportunity to analyse this historical reconstruction of working knowledge. Garfinkel (1967: 204-205), in his study of medical work, makes a similar point when emphasising the productive roles of repetitions and omittance:

> "A subsequent entry may be played off against a former one in such a way that what was known then, now changes complexion. The contents of the folder may jostle each other in bidding to play part in a pending argument. It is an open question whether things said twice are repetitions, or whether the latter has significance, say, of confirming the former. The same hold true of omissions. Indeed, both come to view only in the context of some elected scheme of interpretation"

At the Section of Nephrology this layering is performed during the generation of the discharge letters by enacting – by explicit repetition in the form of textual copying – a particular computational report relevant in the subsequent treatment:

> "We have dialyse-patients that come regularly. For every visit, there are pre-scheduled interventions (…) including some extremely important computations. (...) These computations are important [to us] because they indicate whether the medication and treatment are sufficient. (...) We accordingly paste the result [of these computations] into the discharge letter" (physician, Section of Nephrology)

As a patient spends time in the department, several of the workers get to know the patient through meetings, discussing and assessment of further treatment and informal discussions in the on-duty room. These on-duty rooms are the arenas for collective learning in communities as physicians and nurses regularly come and go, pose questions, discuss cases, share stories and talk on the phone with patients (Brown & Duguid, 1991:46). This transforms individual knowledge into a sort of distributed cognition or a stream of collective self-knowledge recognised by a constant humming with itself about itself (Knorr-Cetina, 1999:173). As part of this, there is an ongoing enactment, refinement and omitting of earlier, historical knowledge representations. The admittance report, the first report filled out during a patient's trajectory, illustrates this as voiced by one physician:

> "The admittance report produced in the emergency department is a mix of previous case history and reason for admittance and what the physician believes is the patient's problem. But if this turns out to be wrong, the admittance report becomes completely useless. What the admitting physician assumed initially, becomes irrelevant both in professional terms and for the receiver of the discharge letter."

The discharge letters contain not only valuable information aimed at informing the general practitioner about what has happened during the stay. It often contains important information essential for the hospital physicians themselves. This is often the case in departments having a lot of regulars such as the Dept. of Oncology:

> "If it is a new patient that will return to the department in order to get radiation treatments or chemotherapy, then we are receivers of the discharge letter as well. The physician who admits the patient the next time relies on the discharge letter to get an overview of prior visits and planned treatments. Thus, we need more details than what is strictly necessary for a general practitioner."

The enactment of the various knowledge representations that go into the generation of the discharge letters is performed in a fluid, interleaved way; the discharge letters are in-the-making. The situation described from the Dept. of Cardiac and Thoracic surgery is illustrative where Pasi gathers necessary information from the pre-visit meeting, like current status of the patient, change in medications and further treatment strategy, which is fed into the dictating process. As we have seen, he also stops dictating for a moment in order to check other information sources in the investigation room, examines the patient by measuring the talus arm index of the patient. He interrupts his dictating in order to ask the head physician for the patient's control strategy.

Not only are the discharge letters moulded by the selective omitting and enacting of existing elements of the total body of knowledge representations, they are also produced with a purpose in mind. They mark the transferral of responsibility from the hospital making the General Practitioners a key recipient group. As a result, the discharge letters, in principle, are generated with a clear sense of what Boland and Tenkasi (1995) describe as perspective taking in mind. A good example of this is the assistant physician in the Dept. of Oncology dictating discharge letters. She underscores that her own ½ year experience as a general practitioner, as a part of their training, has made her aware of what kind of knowledge the general practitioners need:

> "The head physicians possess a lot more experience than us, but maybe we [assistant physicians] are more concerned about what the general practitioners think and pay more attention to its content (…) [and as a former receiver of discharge letters] I try to imagine what kind of information I would have appreciated"

The conclusion part in the discharge letters is extremely important for general practitioners in their follow-up of the patients. Sometimes, however, discharge letters from highly specialised departments are difficult to understand since head physicians in such departments are fairly 'to the point' when dictating:

> "There are discharge letters where the conclusions are difficult to get - whether the patient has as disease or not – particularly from very specialised departments. A specialist can read more between the lines (...) To a general practitioner, it is not clear what the assessment is and how close the patient is in the process towards a diagnose or the current status of the treatment" (general practitioner)

Sense-making is strongly related to the ability to explain and justify. A good example is the lengthy discharge letters in the Dept. of Oncology, like the one whose fragments where presented earlier. The assistant physician underscores that:

> "It is important to think about what the general practitioner need to know. We treat new cancer diseases that is not common in general practice. It is not obvious what to be aware of, like possible side effects of 'cytotoxin'. Things like that [are clear to us but] ought to be included in the discharge letter" (assistant physician, Dept. of Oncology)

The point here is that understanding might increase as redundancy, additional information is available (Czarniawska, 1997:134). The reader is provided with redundant information that might be consulted in the process of getting a better overview. This point comes close to Nonaka and Takeuchi's (1995:230) notion of 'learning by intrusion':

> " 'Learning by intrusion', which means existence of information that goes beyond the immediate operational requirements of each individual. The redundant information enables individuals to invade each other's functional boundaries and offer advice or provide new information from different perspectives"

The resulting discharge letter, then, is not so much a finalised account as a working 'memory' up till this point in time. In producing it, the most recent account is taken as a point of departure but with selectively enacting older, historically buried sediments in the process.

## 5.2. Orchestrating

Knowledge representations about a patient in a hospital are truly dispersed and heterogeneous. There are numerous textual representations (some in paper, some in information systems) accumulated during the patient's trajectory through the different departments and laboratories (including physicians' running notes, nurses' documentation of treatment, laboratory reports, admittance reports and surgery reports) as well as graphical representations (x-ray images, CT- and MR scans and EKG printouts). In addition, there is a rich source of verbalised representations, the ongoing hum of discussing, reporting, joking and story telling of and around patients (Atkinson, 1995; Hunter, 1991). Making sense of the patient thus invariably involves the bringing together of a vast network of heterogeneous knowledge representations. It is essential to recognise how this goes beyond an instrumental 'collection' of these representations. There is a considerable amount of *work* – of a variety of types, more often than not taken for granted and rendered 'invisible' – that goes into the validation, double-checking and sense-making of these representations, a point argued forcefully by Shapin (1994). We study in more detail different mechanisms involved in the rendering of this network of knowledge representations as credible and trustworthy, i.e. useful in everyday practice.

The distributed nature of modern hospital work implies that knowledge about a patient is generated by numerous wards, departments and laboratories. Reading – comparing, contrasting and double-checking – the different accounts is an important way in which knowledge representations are *made* useful. The physician at the Section of Nephrology illustrates this when he double-checked information acquired at the emergency department with information provided by the patient and his wife. Another example is when the physician interleaves his dictation of discharge letters by comparing and contrasting the blood results from the electronic laboratory system with what constitute normal and standard values. At one instance, for instance, the physician has to validate a rather unusual laboratory result for a patient with a ratio between sodium/urine and potassium/urine close to 1 when the standard ratio is 2/1.

Pasi at the Dept. of Cardiology and Thoracic surgery (section 4.2) spends a substantial amount of time (hours) reading the different accounts. He leans heavily on the account by the Dept. of Medicine, which conducted a full examination of the patient before admittance to Pasi's department.

Also, the location of the sources of knowledge enters into the assessment of credibility (Shapin, 1994). This implies that, as part of making sense of a patient, knowledge is validated by involving those 'closer' to the origin of the knowledge as one physician points out:

> "Based on new information in the process [such as laboratory reports, x-rays], we discuss the case with pathologists, radiologists and haematologists as well as internally in our department (...) where the outcome is regularly documented as a note in the patient record".

It is by no mean 'given' what constitute relevant knowledge. It is important that the physician considers the various information sources to check whether they contribute to changes in the case trajectory. A typical example is when both Pasi and the physician at the Section of Nephrology (section 4.3) read the nurses' report to assess whether it contains anything important. As we know, the latter commented that it did not for the patient in question. A similar thing occurred when the physician read and summarised the radiology report. From this report, he selected the information he considered most important as well as performing his own assessments.

Following the same line of argument, new results may implicate that you have to consider the case from a different angle as:

> "Several pieces must be assembled together (...) but for some tests you have to wait a long time. This could delay the assessment of the case, the planning of the patient's follow-up of controls. Sometimes it also has implications for the conclusion of the diagnosis" (physician at the Section of Nephrology)

For the various information sources, work must be invested to evaluate whether they contribute with new knowledge about the case.

Hospitals are notoriously hierarchical institutions. This translates into the issue of legitimising knowledge in the sense that it is highly relevant *who* generates and *where* knowledge comes from (Bowker, Star & Timmermans, 1995; Smith, 1990; Klein & Myers, 1999; Hunter, 1991; Bowker & Star 1999). For our purposes, the relevant aspect is that this spawns strategies and mechanisms – in short, work – to render the knowledge credible.

The credibility of a given piece of data in the patient's chart, an advice, or diverging diagnosis, is tightly linked to who's observation or opinion it was (Atkinson, 1995:57). This draws an important distinction between experts and non-experts (Knorr-Cetina 1999:131).

This is illustrated by the difference between head physicians as experts and assistant doctors as non-experts. A head physician is assumed to possess high-level competence within the actual special field, while the assistant can possess various degree of competence as one assistant physician explains:

> "If it is some of the regulars that have produced the summary then I might accept it as 'face value' and use it as it is. But if it is produced by an inexperienced physician I have to read more thoroughly to check whether it can be used"

Similarly, the work of assistant physicians (some just out of medical school) may be rendered worthless as one chief physician comments with reference to the admittance reports:

> "Often it is the novices who receive the patients. [Which means] that the assessment can be turned around the following day making the summary worthless"

This reiterates Shapin's (1994:38) point about the deeply social character of knowledge in the sense that:

> "Insofar as knowledge comes to us via other people's relations, taking in that knowledge, rejecting it, or holding judgment in abeyance involves knowledge of *who these people are.* … What of relevance to credibility assessment do we know about them as individuals and as members of some collectivity?"

The amount of additional work in rendering knowledge credible is of course not constant. It depends on the level of credibility already ascribed to a given source. Credibility is an acquired quality, it is not postulated. This is illustrated by patients that are hospitalised for a relatively long time as a part of their treatment:

> "Gradually, we get to know the patients very well and [the nature of the disease] involves them a lot and they must agree on what we decide to do" (Physician, Dept. of Oncology)

Summing up, the distributed nature of large, modern hospital generates knowledge representations of a wide range of types of origins. Producing a discharge letter requires obtaining an overview of these. This goes well beyond the mere 'collecting' of the information. Knowledge is rendered credible through (largely invisible) *work* by a range of mechanisms including reading, interpreting, assessing, comparing, contrasting and negotiating.

## 5.3. Organising

Hospitals, as sites of knowledge work, are large, complex and dynamic organisational entities. The complexity has several sources. There are a large number of distinct health professions with associated communities of practice and with different political standing in the hierarchy. The collection of tools, artefacts and equipment is significant. This spans from a variety of utterly mundane artefacts such as report templates and archives to high-tech equipment like MR scanners requiring competent and specialised users. The trajectory of a patient during a stay spawns a comprehensive set of work tasks. The organisation – the coordination, delegation, tracking and accumulation – of this is not viable through centralised, hierarchical control. The result, then, is that this organising is performed *as part of* the ongoing production of knowledge representations. This supplements the predominantly cognitive bias the discourse on knowledge work with its emphasis on innovation, creativity and flexibility (Blackler, 1995; Nonaka & Takeuchi, 1995; Alvesson, 2001). As Berg (1999:388, following Schmidt and Simone) points out:

> "Reading and writing artefacts coordinate activities, then, through a 'precomputation of task interdependencies' which 'reduces the space of possibilities' for the entities that interrelate with it"

reiterating Smith (1990:217-218) holding that:

> "the organised character of formal organisation, depends heavily on textual practices, which co-ordinate, order, provide continuity, monitor, and organise relations between different segments and phases of organisational courses of action"

Based on these insights we analyse in more detail how the production of discharge letters play an important role in organising work. More specifically, we study (i) how the artefact (discharge letter) is used as a token for synchronisation and (ii) how the practice of completing it feeds into the scheduling of future activities. The discharge letter is pivotal in the division of labour between the hospital and general practitioners. As there might be overlapping responsibilities between the hospital and the general practitioner, establishing who is responsible for what is essential as one general practitioner argues:

> "It is vital for us is to have established distribution of further responsibility, [that is], what is the responsibility of the hospital and what is the responsibility of the general practitioner"

This is mirrored in a chief physician's (Dept. of Ear, Nose and Throat, section 4.1) emphasis that:

> "when the discharges letters are specific and brief they might serve as manageable instructions from the hospital to the general practitioner"

The discharge letter also coordinates activities between hospitals as is the case when Pasi (Dept. of Cardiac and Thoracic Surgery, section 4.2) has to produce a discharge letter before the patient is allowed to leave. In fact several people in the department are involved in this effort. The discharge letter serves as an instruction for the follow-up procedure in the local hospital. This information is rather important and as one of the secretaries tells:

> "There have been instances where the air-ambulance people has refused to take-off as a result of lacking discharge letters"

Prior to his surgery, this patient, as a standard procedure has had a full examination in the Dept. of Medicine. This means that at this department the physicians have evaluated different treatment strategies and concluded that the conditions for surgery were fulfilled. Thus the baton was handed over to the Dept. of Cardiac and Thoracic Surgery. This occurred among other things through an extensive discharge letter from the Dept. of Medicine.

Yet another example of coordination is the PD-form for dialyse patients shown earlier in figure 1. This form is not merely an externalised knowledge representation of the state of a patient. It simultaneously acts as a vehicle for the delegation and coordination of the required work tasks. The filling in of the cells in the form is delegated across the different professional groups and persons. The degree of completion of the form represents an account of what has been done up till a certain point in time, hence also what remains (Berg, 1996).

For instance, the PD-form inscribes instructions about actions to the nurses in the section as illustrated by the physician's reading of the form:

> Firstly, here [pointing to the 'Next control' field in the PD-form] it is stated that the patient will be summoned to the next control in the Medical policlinic, with whom, and 2 months from now. Secondly, 'New PD-control with PET in May 2002' (...) The head nurse will get a copy of the discharge letter and will summon the patient to the next control. The nurses are skilled in running the standard procedures. What is special, hence highlighted in the PD-form, is that for this patient's next visit we take specimens from the dialyse-solution in order to perform a PET analysis"

In this way, the PD-form feeds into the work itself and is a part of it or as Berg (1996:9) underscores:

> "It does not merely represent this coordination of work: it stipulates and mediates it. It is a material form of semi-public memory: relieving medical personnel's burden of organising and keeping track of the work to be done and its outcomes."

In fact, the PD-form gives a brief and accumulated overview of planned activities for the patient, which according to Berg (1999:391) affords "an increase in complexity of the work practice without a simultaneous increase in complexity in individual interactions".

Sometimes the discharge letter may explicitly serve as an instruction to the physicians themselves as is clear when the standard procedure in the Section of Nephrology is to read the discharge letter associated with the previous stay (see earlier section). Often, results from the laboratory are not ready until the patient has left:

> "The PD-patients are hospitalised for 2 days during which some laboratory tests are carried out. However you cannot initiate the medical procedures until the laboratory results are ready [in the meantime the patient has left] and then you have to initiate them next time. Therefore the discharge letters for PD-patient contains a conclusion instructing what to the next time around"

Supplementing the cognitive aspects of knowledge work, knowledge representations (such as the perfectly mundane discharge letter) perform essential tasks in regulating, coordinating and controlling the organisation of work both within and outside the hospital.

## 6. Conclusion

Information, one of the many knowledge representations, abounds in organisations. This spawns perfectly understandable desires for tighter integration of information systems. The dominant approach which favours centralised, monolithic solutions – illustrated by the recent attention devoted to Enterprise Resource Planning (ERP) systems – is problematic given our analysis. Information needs to be *rendered* credible and trustworthy to be useful in everyday work (Hasselbring, 2000; Ellingsen & Monteiro, 2002). It is the (largely invisible) work involved in this rendering that makes it credible (Shapin, 1994). Integration in the sense of bringing together or 'collecting' information fails to reproduce these mechanisms of rendering knowledge credible. For instance, a dichotomous distinction between explicit (codified) and tacit knowledge is accordingly a simplification that glosses over the issue of exactly *how*, in the fluid interleaving of codified and narrative forms, knowledge becomes credible. Similarly, for settings like the one we have studied, the historicity of knowledge is an intrinsic aspect. This implies that knowledge needs to be enacted – repeated, narrated, kept vivid at the expense of other – to be relevant for a given purpose at hand (such as writing of discharge letters).

Ethnographically inspired accounts of the use of information systems might be misconstrued as an (implicit !) argument of preserving status quo as existing practice is portrayed as a contingent, delicate balance too fragile to be touched. It would, however, be frustrating for practitioners and disappointing for IS researchers to deny the possibility of transformations, of changes to the contents and organisation of knowledge work (Berg, 1999).

Given, then, that changing practice is indeed possible, what implications about knowledge work can be drawn from our analysis of clinical practices in producing and consuming discharge letters in large hospitals ?

Our analysis points out how core activities in the actual *rendering* of making knowledge useful are not supported. The dominant position still in supporting knowledge work with information systems tends to assume that the type of work that go into this rendering is non-existent or automatic (Zack, 1999). In this sense, supporting knowledge work with information systems needs to take seriously the finer socio-technical mechanisms we have identified.

Similarly, the discourse on the integration of information systems focuses almost exclusively (Hasselbring, 2000; see Ellingsen & Monteiro, 2002) on how integration amounts to gaining *access* to relevant information (e.g. 'clickable' access to reports and notes). Again, this assumes too much. It assumes that the rendering of the information as working knowledge somehow happens. But as we have shown, it is work rather than mystery that makes this happen. The question, at the present very much an open one, is exactly how.

Activities such as reading, interpreting and validating of information is difficult to support in any direct manner through information systems. Yet, given an increased acknowledgment of these largely 'invisible' activities of knowledge work, information systems could be used to facilitate keeping an *overview* of potentially relevant information (e.g. highlighting historical summaries and examinations) and keeping *track* of what kind of information has been used (e.g. checking off from a list). This would imply using information systems to highlight, rather than neglect, the sources of the information, the version/ date and the state of (in)completion of an interpretation.

## Acknowledgment

This work has in part been sponsored by Norwegian Research Council. We are grateful to ongoing discussions and comments from the Kvalis project group (http://kvalis.ntnu.no). Thanks to Knut Hansen at UNN who has gave us direct feedback on earlier drafts on the paper. Our anonymous reviewers also provided valuable comments, especially regarding the work of Shapin